\documentclass{aa}
\usepackage{hyperref}
\usepackage{txfonts}
\usepackage{graphicx} 
\usepackage{psfig,longtable,lscape}
\usepackage{epsfig}
\usepackage[ ]{natbib} 
\usepackage{multirow}
\begin{document}

\title{Discovery of periodic dips in the light curve of GX 13+1: the X-ray  orbital ephemeris of the source}

\author{R. Iaria\inst{1},
T. Di Salvo\inst{1}, L. Burderi\inst{2},  A. Riggio\inst{2},
 A. D'A\`\i\inst{1}, N. R. Robba\inst{1}}

\offprints{R. Iaria, \email{rosario.iaria@unipa.it}}

\institute{Dipartimento di Fisica e Chimica,
Universit\`a di Palermo, via Archirafi 36 - 90123 Palermo, Italy
\and
Dipartimento di Fisica, Universit\`a degli Studi di Cagliari, SP
Monserrato-Sestu, KM 0.7, Monserrato, 09042 Italy}

\date{\today}

\abstract
{The bright low-mass X-ray binary (LMXB) GX 13+1 is one of the most
  peculiar Galactic binary systems. A periodicity of 24.27 d with a
  formal statistical error of 0.03 d was observed in its power
  spectrum density obtained with RXTE All Sky Monitor (ASM) data
  spanning 14 years.}
{Starting from a recent study, indicating GX 13+1 as a possible
  dipping source candidate, we systematically searched for periodic
  dips in the X-ray light curves of GX 13+1 from 1996 up to 2013 using
  RXTE/ASM, and MAXI data to determine for the first time the X-ray
  orbital ephemeris of GX 13+1.}
{We searched for a periodic signal in the ASM and MAXI light curves,
  finding a common periodicity  of 24.53 d. We folded the 1.3-5 keV and
  5-12.1 keV ASM light curves and the 2-4  and 4-10  keV MAXI light
  curves at  the period of 24.53 d finding a
  periodic dip.  To refine the value of the period
  we used the timing technique dividing the ASM light curve in eight
  intervals and the MAXI light curve in two intervals, obtaining four
  and two dip arrival times from the ASM and MAXI light curves,
  respectively.}
{We improved the X-ray position of GX 13+1 using a recent Chandra
    observation. The new  X-ray  position  is discrepant by $\sim 7\arcsec$ 
  from the previous one, while it is compatible with the infrared and radio 
    counterpart positions. We detected an  X-ray dip, that is
  totally covered by the Chandra observation, in the light curve of
  GX 13+1 and showed, a-posteriori, that it is a periodic dip.  
We obtained seven dip
  arrival times from ASM, MAXI, and Chandra light curves.  We
  calculated the delays of the detected dip arrival times with respect
  to the expected times for a 24.52 d periodicity.  Fitting the delays
  with a linear function we find that the orbital period and the epoch
  of reference of GX 13+1 are 24.5274(2) days and 50,086.79(3) MJD,
  respectively. Adopting a quadratic ephemeris we do not improve the
  fit. The inferred orbital period derivative of $ 8(37) \times 10^{-8}$ s/s,
  with the error at 68\% confidence level, does not allow us to constrain 
 the orbital evolution of the binary system.}
{We demonstrated the existence of periodic dips in the ASM and MAXI
  light curves estimating that the orbital period of GX 13+1 is
  24.5274(2) d. GX 13+1 has the longest known orbital period for a
  Galactic neutron star LMXB powered by Roche lobe overflow.}
\authorrunning {R.\ Iaria et al.}
\titlerunning {The first ephemeris of GX 13+1}

\maketitle

\section{Introduction}
  \begin{figure*}
\includegraphics[height=4.9cm, angle=0]{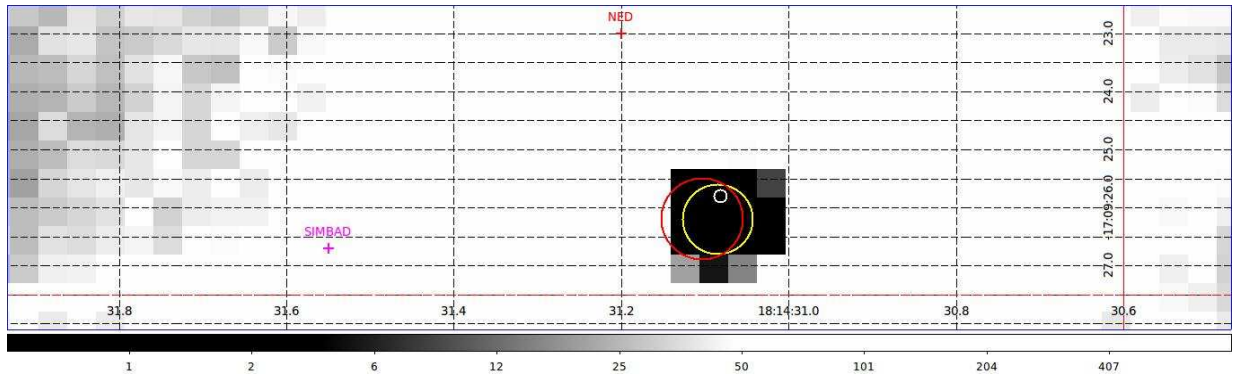}
\caption{Chandra/HETGS  image of GX 13+1 obtained from the reprocessed
    level2-event file. The image is a detail close to the X-ray
    position. The grey scale represents the number of photons for each
    pixel integrated along the whole duration of the
    observation. Close to the zero-order position the number of events
    is almost zero because of the strong pile-up. The yellow circle centred at
  RA$=273\degr.62952$ and DEC$=-17\degr.157275$ indicates the updated
  position of GX 13+1 reported in this paper, the red and white
  circles indicate the error boxes of the infrared and radio
  counterparts of GX 13+1. The magenta and the red crosses indicate
  the previous X-ray position of GX 13+1 in the SIMBAD and NED
  catalogue, respectively.}
\label{Figpos}
\end{figure*}
  \begin{figure*}
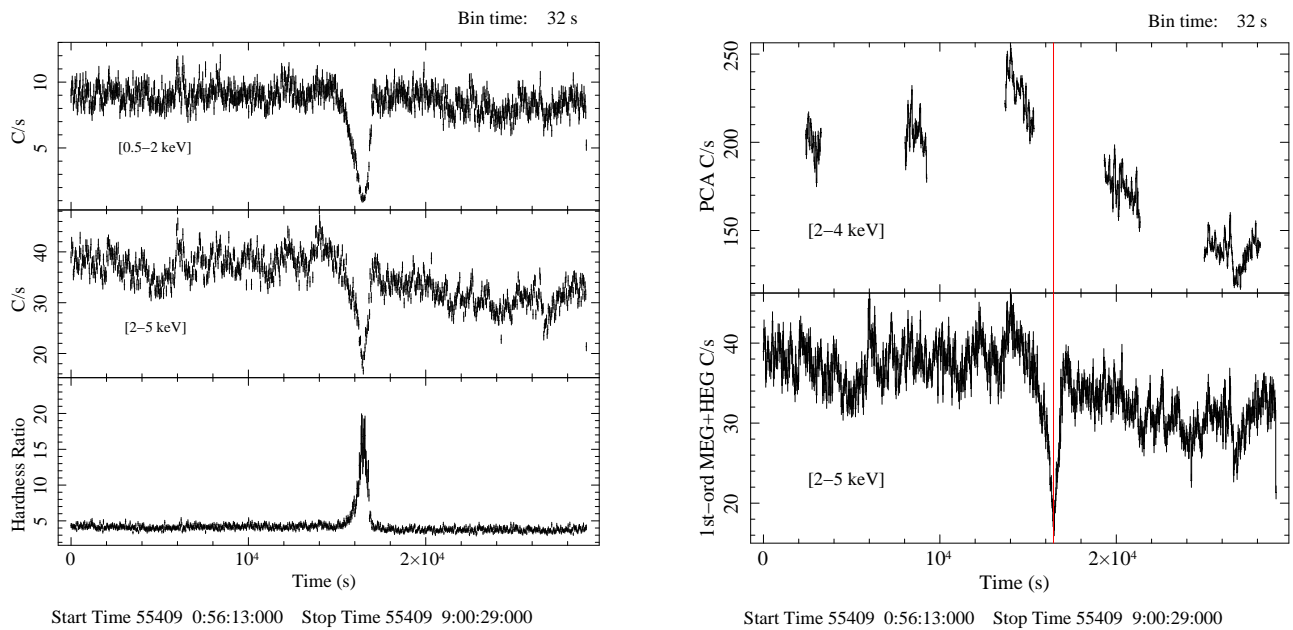

\includegraphics[angle=0, width=\columnwidth]{hardness_bary_chandra.ps}
\includegraphics[angle=0, width=\columnwidth]{PCA_CHANDRA_REF.ps}
\caption{ Left Panel: Chandra light curve of GX 13+1 between 
55,409.03 and 55,409.37
  MJD.  Summed first-order MEG and HEG light curves in the 0.5-2 keV
  energy band (top), in the 2-5 keV energy band (middle),
  and the corresponding hardness ratio (bottom) are shown. The bin time is
  32 s. Right Panel: background subtracted RXTE/PCA light curve of GX 13+1 in the 2-4
  keV energy band (top) and Chandra light curve of GX 13+1 in the 2-5
  keV energy band (bottom). The solid vertical line indicates the estimated dip
  arrival time.  The bin time is 32 s.}
\label{Fig1}
\end{figure*}
GX 13+1 is a bright low-mass X-ray binary (LMXB); the detection of
sporadic X-ray bursts \citep{Fleisc_85,Matsuba_95} suggests that the
accreting compact object is a neutron star.  Using infrared data
\cite{Band_99} estimated that the spectral type of the companion star
is K5 III and the distance to the source is $7 \pm 1$ kpc.
\cite{Ueda_01} detected,  for the first time in a LMXB containing a
neutron star, an absorption line, associated with 
\ion{Fe}{xxvi} K$\alpha$, in the ASCA/SIS X-ray spectrum of GX 13+1,
suggesting that the inclination angle of the system is relatively
large. Using XMM/Epic-pn data \cite{Sidoli_02} found a more complex
structure in the Fe-K region detecting absorption lines associated
to \ion{Ca}{xx}, \ion{Fe}{xxv} K$\alpha$ and \ion{Fe}{xxv} K$\beta$,
\ion{Fe}{xxvi} K$\alpha$ and \ion{Fe}{xxvi} K$\beta$ transitions.
\cite{Ueda_04} studied the high-energy resolution spectrum of GX
  13+1 taken by the Chandra high energy transmission grating
  spectrometer (HETGS) observing several absorption lines from highly
  ionised ions of Fe, Mn, Cr, Ca, Ar, S, Si, and Mg. The authors
  detected a significant blueshift of those lines suggesting that the
  ionised plasma has an outflow velocity of $\sim 400$ km s$^{-1}$.
  Furthermore, \cite{Ueda_04} inferred a mass outflow rate of
  $\dot{M} \geq 7 \times 10^{18} [(\Omega_{tot}/4 \pi)/0.4]$ g
  s$^{-1}$, where $\Omega_{tot}$ is the total solid angle subtended by
  the plasma; this is comparable to the mass accretion rate in the
  inner part of the disk, $10^{18}$ g s$^{-1}$, estimated from the
  continuum spectrum.

Recently \cite{Trigo_2012} analysed several XMM-Newton data sets of GX
13+1 fitting the continuum emission with a model consisting of
disc-black body plus a black body component together with a Gaussian
emission line at 6.55--6.7 keV. The authors also  found that the
continuum emission is absorbed by cold and photo-ionised material and
constrained the inclination of GX13+1 to be between 60 and
80$^{\circ}$ because of the presence of strong absorption  along
the line of sight, obscuring up to 80\% of the total emission in one
observation, and the absence of eclipses.
 Although GX 13+1 is thought to be a high inclination angle system
 belonging to the Dipping source class of LMXBs the estimation of its
 orbital period is not simple because it is quite large.

 Analysing the power spectrum of GX 13+1 obtained from RXTE All-Sky
 Monitor (ASM) data spanning 7 years \cite{Corbet_03} detected a
 period of $24.07 \pm 0.02$ days. \cite{Corbet_2010}, using RXTE ASM
 data spanning 14 years, estimated a period of $24.27 \pm 0.03$ days.
 If the periodicity of 24 days is the orbital period of the binary
 system then GX 13+1 has the longest known orbital period for a
 Galactic neutron star LMXB; however,
 \cite{Corbet_2010} noted that the periodic signal observed in the
 power spectrum is not strictly coherent and suggested that it might
 be caused by a structure that is not completely phase-locked with the
 orbital period.  \cite{Corbet_2010} proposed that the X-ray
 variability may be caused by an unresolved dipping modulation.

 In this work we report the first ephemeris of GX 13+1 with an
 accuracy of the orbital period by a factor one hundred
 larger than the previous value proposed by \cite{Corbet_2010}. The
 periodic dips are clearly observed in the RXTE/ASM light curve
 spanning almost 16 years, in the MAXI light curve spanning 4 years,
 and during a recent Chandra/HETGS observation.

\section{Observations}

\subsection{Chandra}   
Chandra observed GX 13+1 nine times from 2000 to 2011. The
corresponding obsids are 950 \citep[see][]{Smith_02}, 2708
\citep[see][]{Ueda_01}, 6093 \citep[see][]{Smith_08}, 11814, 11815,
11816, 11817, 11818, and 13197.  We searched for the presence of dips
in these light curves, finding one clear dip signature in the obsid
11814.  It was taken on 2010 August 1 00:31:31 UT for an exposure time
of 29 ks and performed in Timed Graded mode using the High Energy
Transmission Grating Spectrometer (HETGS). We reprocessed the
  data and used the level2-event file to update the X-ray coordinates of
  GX 13+1. Because the zero-order X-ray image is strongly piled-up, the
  new coordinates were determined by fitting the intersection of the
  grating arms positions with the source readout streak (similarly to
  the pipeline adopted by \cite{Iaria_06} and \cite{Iaria_07}). 
 The updated X-ray coordinates of the source are
  RA$=273\degr.62952$ and DEC$=-17\degr.157275$ (referred to J2000.0)
  with an associated error of $0\farcs6$, that is the overall 90\%
 uncertainty circle of the Chandra
  X-ray absolute positional accuracy\footnote{See 
\url{http://cxc.harvard.edu/cal/ASPECT/celmon/} for more
    details}. We obtained very similar values
using the Chandra script {\it tg\_findzo}.
The new X-ray position  is at
 $6\farcs7$ from that reported in the SIMBAD catalogue,
 while it is compatible with the infrared
\citep{Garcia_92} and radio  \citep{Grindlay_86}  counterpart
positions. We show in Fig. \ref{Figpos} the new estimated position
with the corresponding error marked by a yellow circle; the red and
 white circles indicate the error boxes of the infrared and radio
counterpart positions of GX 13+1.
The  magenta and  red crosses indicate the previous X-ray position of
GX 13+1 in the SIMBAD and NED catalogue, respectively.
 
We barycentred the level-2 event file with
respect to the new X-ray coordinates using the tool {\it axbary} in
the CIAO package, and, finally, extracted the light curves
in the 0.5-2 keV and 2-5 keV energy bands from the level-2 event file
 using the Chandra tool {\it dmextract} adding the first-order MEG
and HEG and adopting a bin time of 32 s.  The light curves and the
corresponding hardness ratio (HR) are shown in Fig. \ref{Fig1} (left
panel).
  \begin{figure}
\includegraphics[angle=0, width=\columnwidth]{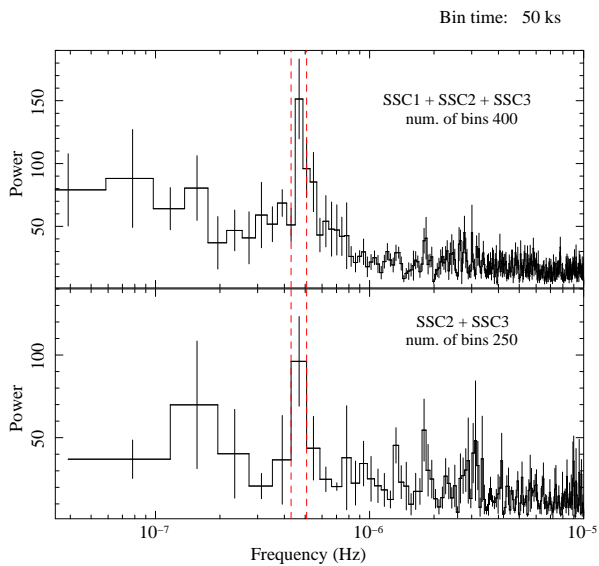}
\caption{ Power spectrum densities of the 
1.3-12.1 keV ASM light curve including all the SSCs (top panel) and only SSC2 and SSC3 (bottom panel).
We adopted a bin time  of 50 ks for both the light curves and  a number of bins of 400 and 250, respectively.
A periodic signal is evident in both the light curves and corresponds to a period between 22.78 
and 26.92 d that are indicated with the vertical dashed-red lines.}
\label{pow_asm}
\end{figure}

The large increase of the HR around 16,000 s from the beginning of the
observation (see left-bottom panel of Fig.  \ref{Fig1}) suggests a
large photoelectric absorption; the 0.5-2 keV count rate decreases
more rapidly than the 2-5 keV count rate, that is typical of a dip
phenomenon.  To estimate the arrival time of the dip we fitted the HR
data between 16,000 and 16,900 s, corresponding to the maximum of the
hardness ratio, with a constant plus a Gaussian profile assuming the
centre of the Gaussian to be the arrival time of the dip.  We assumed
as a 1-sigma uncertainty to this measure the statistical error derived
from the best-fitting centre of the Gaussian.  The observation was
taken between 55,409.03 and 55,409.37 MJD.  We estimated that the
minimum of the dip occurs at 55,409.22935(11) MJD.

Simultaneously the PCA on board of {\it Rossi-XTE} (RXTE) observed GX
13+1 on 2010 August 1 from 01:29:03.56 UT to 08:38:55.56 UT
(observations P95338-01-01-04, P95338-01-01-03, P95338-01-01-02,
P95338-01-01-01, and P95338-01-01-05).  We show the RXTE/PCA 
background-subtracted light
curve of GX 13+1, obtained from the standard products in the 2-4 keV
energy range, in the right-top  panel of Fig. \ref{Fig1}.  For sake of
clarity we also show in the right-bottom panel of Fig. \ref{Fig1} the summed
first-order MEG and HEG light curve in the 2-5 keV energy range.
 Unfortunately, the RXTE/PCA observation does not cover the dip observed 
by Chandra. The dip arrival time is indicated with the solid vertical line in
Fig.   \ref{Fig1} (right panels).  

\subsection{RXTE/ASM and MAXI}
\label{pippo}
 \begin{figure}
\includegraphics[angle=0, width=\columnwidth]{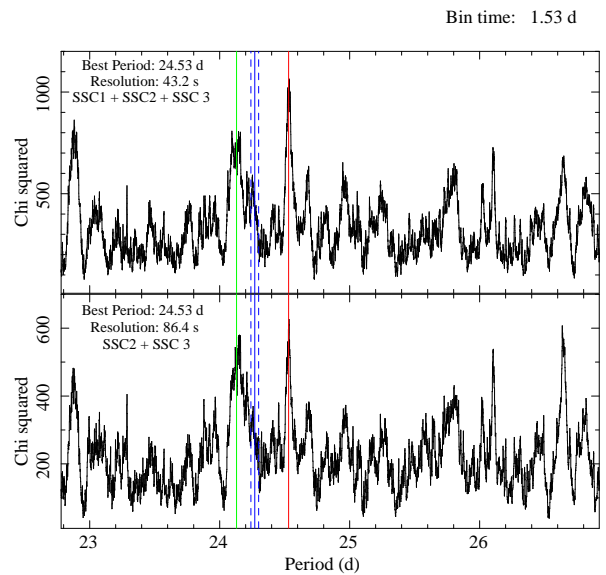}
\caption{Folding search of periodicities between 22.78 and 26.92 d in
  the ASM light curve including all the SSCs (top panel) and
 including only SSC 2 and 3 (bottom panel), respectively. We adopted 
    16 phase-bins per period for the trial folded light curves and a
    resolution in the period search of 43.2 s (top panel) and 86.4 s
    (bottom panel).  The maximum of $\chi^2$ is observed at 24.53 d
  (red vertical line) in both the curves.  The vertical green-solid
  line indicates a possible periodicity at 24.12 d.  The vertical
  blue-solid and blue-dashed lines indicate the period of $24.27 \pm
  0.03$ d suggested by \cite{Corbet_2010}.}
\label{win_2}
\end{figure}

With the aim to find periodic dips, we analysed the light curve of GX
13+1 taken by the ASM on board RXTE; the light curve spans around 16
years from 50,088.21 to 55,865.44 MJD.  The ASM consists of three
Scanning Shadow Cameras (SSCs) mounted on a rotating drive assembly.
First of all, we barycentred the ASM light curve of GX 13+1 with
  respect to the new X-ray coordinates using the tool {\it
  faxbary}.  We extracted the 1.3-12 keV ASM light curve from each
SSC. The SSC 1 light curve shows a slightly decline of the flux 
 in agreement with what is
reported by \cite{Corbet_2010}. Using the ftools {\it powspec} in the
XRONOS package we produced two power spectrum densities (PSDs).
  The first PSD was obtained from the ASM light curve including all
  the SSC events (hereafter SSC123 light curve) and the second one
  including only the SSC2 and SSC3 events (hereafter SSC23 light
  curve). We used a bin time of 50 ks for both the PSDs and a number
  of bins of 400 and 250 for the SSC123 and SSC23 light curve,
  respectively.  A periodic signal between 22.78 and 26.92 d
  (indicated in frequency with the vertical red-dashed lines in
  Fig. \ref{pow_asm}) is evident in both the PSDs.  We explored the
period-window between 22.78 and 26.92 d using the ftools {\it
  efsearch} in the XRONOS package for both the light curves.  We
  adopted 16 phase-bins per period  (that is a bin time close to
  1.53 d) for the trial folded light curves and a resolution of the
  period search of 43.2 s and 86.4 s for the SSC123 and SSC23 light
  curve, respectively.  We observe the largest peak of $\chi^2$ close
to 24.53 d for both the light curves (vertical red line in
Fig. \ref{win_2}), a smaller peak is observed at 24.12 d that is indicated
in Fig. \ref{win_2} with a vertical green-solid line.  We also note
that the periodicity of $24.27 \pm 0.03$ d (vertical blue-solid
  and blue-dashed lines in Fig. \ref{win_2}) suggested by
\cite{Corbet_2010} is not significantly detected in our search.  
Since we obtained the
same periodic signals independently of the inclusion or exclusion of
the SSC1 events, we conclude that the decline of the flux in the SSC1 does not affect our analysis. We, therefore, used, in the following,
 the ASM light curves including
all the SSCs to obtain the largest available statistics.
  \begin{figure}
\includegraphics[angle=0, width=\columnwidth]{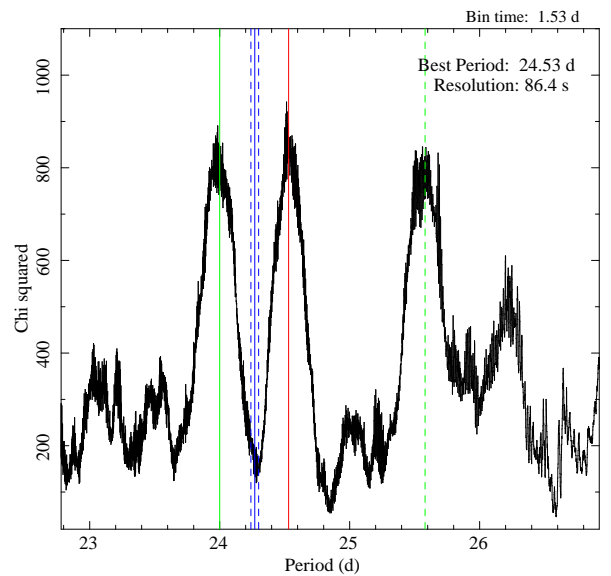}
\caption{Folding search of periodicities in the 2-10 keV MAXI light
  curve between 22.78 and 26.92 d. We adopted 16 phase-bins per period  
for the trial folded light curves and a resolution of period search of 
  0.001 d. The vertical blue-solid and blue-dashed lines
  correspond to the period  $24.27 \pm 0.03$ d suggested by
  \cite{Corbet_2010}. Three peaks of $\chi^2$ are detected: at 24.53 d
  (vertical red line), at 25.6 d (vertical green-dashed line), and at
   24 d (vertical green-solid line).  The latter period is an
  instrumental artifact (see text).}
\label{maxi_efsearch}
\end{figure}
  \begin{figure}
\includegraphics[angle=0, width=\columnwidth]{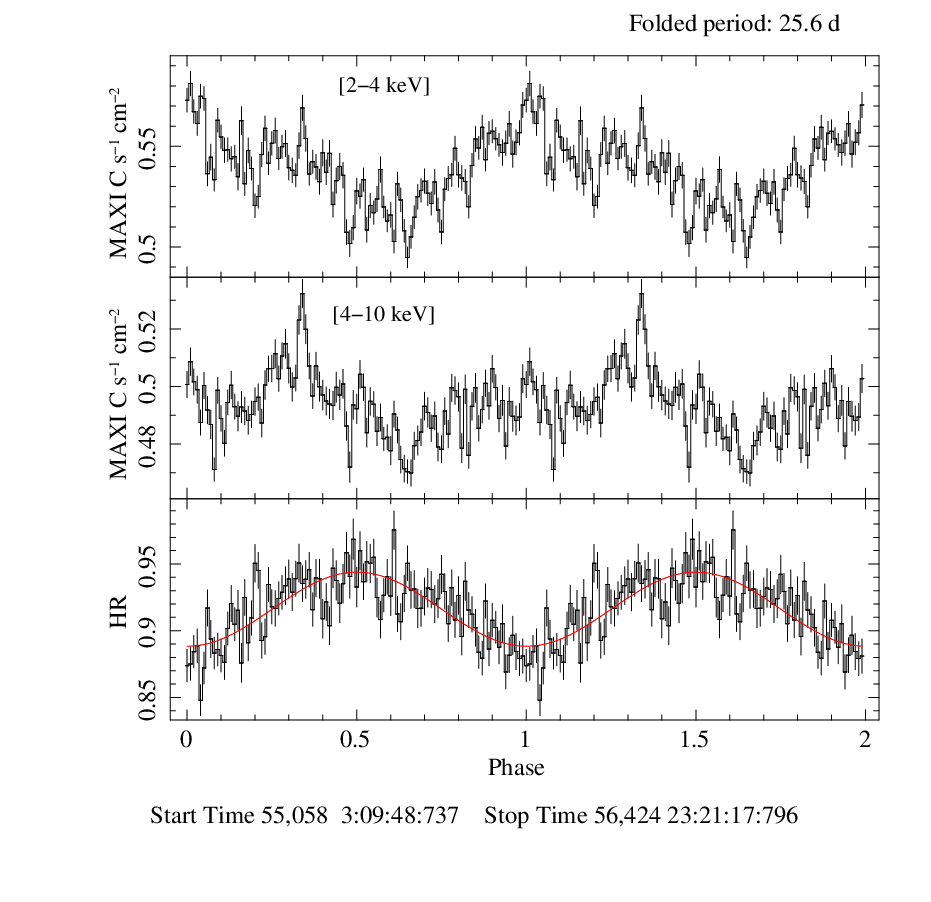}
\caption{ 25.6 d folded 2-4 keV and 4-10 keV MAXI light curves (upper
  and middle panel, respectively).  The HR is shown in the bottom
  panel. We adopted 100 phase-bins per period and Epoch 55,044.77
  MJD. The HR shows an evident sinusoidal modulation (red curve in
  bottom panel) at the folded period.}
\label{maxi_25d6}
\end{figure}

We searched for periodic signals also in the one-orbit 2-10 keV light
curve of GX 13+1 taken by the Monitor of All-sky X-ray Image (MAXI) on
board of the International Space Station (ISS)
\citep[see][]{Matsuoka_09}.  The light curve spans about 4 years.
Adopting the new X-ray coordinates of GX 13+1, we barycentred the
light curve with respect to the centre of mass of the solar system
using the ftool {\it earth2sun}.  Using the ftool {\it efsearch}
  we searched for periodicities between 22.78 and 26.92 d adopting 16
  phase-bins per period in the trial folded light curves and a
  resolution of the period search of 86.4 s.  We found three peaks of
the $\chi^2$ which were fitted with Gaussian profiles; the centres of
the three Gaussian profiles  are close to 24 d, 24.53 d and 25.6 d
(see Fig.  \ref{maxi_efsearch}).  The periodicity at 24.53 d (red-solid
line in Fig.  \ref{maxi_efsearch}) is the same found analysing the ASM
light curve.  The periodic signal at 24 d (green-solid line in Fig.
\ref{maxi_efsearch}) is an instrumental signal due to a not
appropriately subtraction of the background from the data collected by
the 1550V cameras. The 1550V cameras collect events from the region
of sky containing GX 13+1 for 24 days with respect to the total period
of precession of the ISS (i.e. 70 days) and it leaves an excess of
counts which varies in 24 days (Tatehiro Mihara, MAXI team; private
communication).  The periodicity at 25.6 d (green-dashed line in Fig.
\ref{maxi_efsearch}) is similar to the period of $25.8 \pm 0.03$ d
observed in the K-band light curve of GX 13+1 \citep{Corbet_2010}. A
similar but less significant periodic signal seems to be present in
the ASM light curve at $\sim$25.8 d.  We plot the 25.6 d folded 2-4
and 4-10 keV MAXI light curves and the corresponding HR in Fig.
\ref{maxi_25d6} using as epoch 55,044.77 MJD and 100 phase-bins per
period.  A clear sinusoidal modulation with a period of 25.6 d is
present in the HR. However, because we are interested in the presence
of possible periodic dips in the ASM and MAXI light curves we will not
discuss further this periodic signal.  Finally, the vertical
blue-dashed and blue-solid lines indicates the period of $24.27 \pm
0.03$ d suggested by \cite{Corbet_2010}, that is not significantly 
detected in our search.

\begin{figure*}
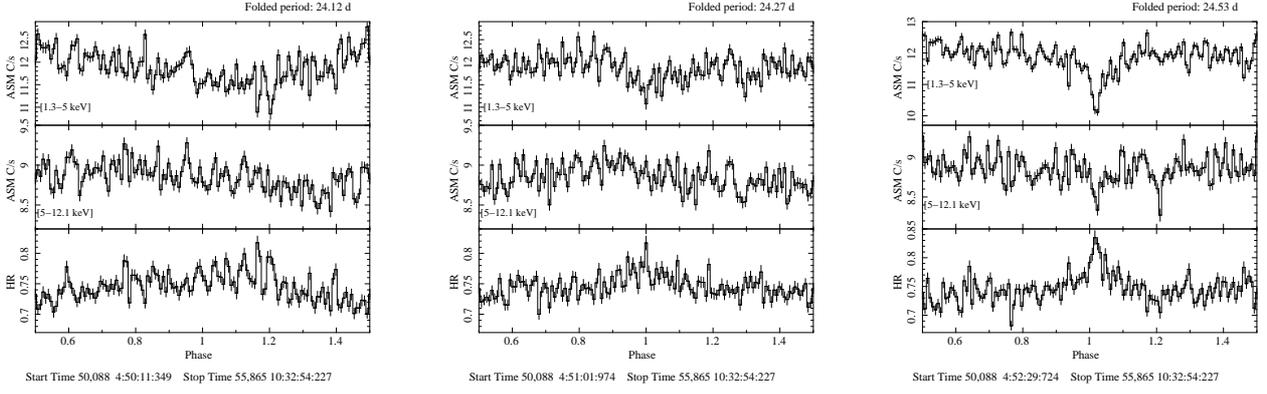

\includegraphics[height=5.6cm,angle=0]{folde_soft_hard_period_24d12.ps}
\includegraphics[height=5.6cm,angle=0]{folde_soft_hard_period_24d27.ps}
\includegraphics[height=5.6cm,angle=0]{folde_soft_hard_period_24d53.ps}
\caption{RXTE/ASM folded light curves of GX 13+1 using the folding
  period 24.12 d (left panel), 24.27 d (middle panel), and 24.53 d
  (right panel).  We show the folded light curves in the 1.3-5 keV
  (upper panels) and 5-12.1 keV (middle panels) energy band, and the
  corresponding hardness ratios (bottom panels) for each folding
  period.  The folded light curves are obtained using 128 phase-bins
  per period.}
\label{Fig2}
\end{figure*}

We folded the ASM light curves in the 1.3-5 keV and 5-12.1 keV using
as folding period 24.12 d, 24.53 d, and 24.27 d, respectively,
  obtaining three folded light curves for each adopted energy band.
The ASM folded light curves and the corresponding HRs are shown in
Fig. \ref{Fig2}; the folded light curves are obtained using 128 phase
bins per period.  The 24.53 d folded light curves highlight the
presence of a dip in the ASM light curve of GX 13+1. The 1.3-5 keV and
5-12.1 keV light curves show a decrease of the count rate from 12.2 to
10.1 c/s and from 8.9 to 8.3 c/s, respectively, close to
phase 0 (see Fig.  \ref{Fig2}, right panel). At the same phase the HR
increases from 0.75 to 0.84.  This suggests that the equivalent
hydrogen column density associated to the local neutral (or partially
ionised) matter increases close to phase 0 causing a reduction of the
count rare at low energies and, consequently, an increase in the HR.
A similar modulation is not observed at  24.12 d (see
Fig. \ref{Fig2}, left panels) and it is less significant using as
folding period 24.27 d (see Fig. \ref{Fig2}, centre panels).
\begin{figure*}
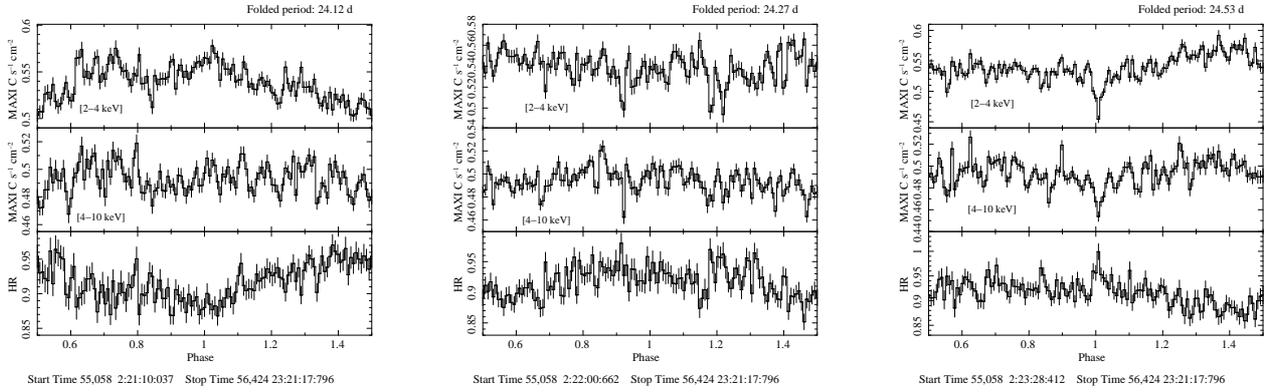

\includegraphics[height=5.6cm,angle=0]{folded_soft_hard_24d12_binned128.ps}
\includegraphics[height=5.6cm,angle=0]{folded_soft_hard_24d27_binned128.ps}
\includegraphics[height=5.6cm,angle=0]{folded_soft_hard_24d53_binned128.ps}
\caption{MAXI folded light curves of GX 13+1 using as folding period
  24.12 d (left panel), 24.27 d (middle panel), and  24.53 d (right
  panel).  We show the folded light curves in the 2-4 keV and 4-10 keV
  energy band and the corresponding hardness  ratios for each folding
  period.  The folded light curves are obtained using 128 phase-bins 
per period.}
\label{Fig2b}
\end{figure*}
We show the 2-4 and 4-10 keV MAXI folded light curves and the
corresponding HRs in Fig. \ref{Fig2b}. We folded them using the
periods of 24.12, 24.27 and 24.53 d and adopting 128 phase-bins per
period.  Also in this case we found that a periodic dip is present
close to phase 0 in the 24.53 d folded MAXI light curves
  (Fig. \ref{Fig2b}, right panels); the dip is absent in the 24.12 and
  24.27 d folded MAXI light curves.  During the dip the MAXI count
rate per cm$^{-2}$ decreases from 0.54 to 0.46 and from 0.49 to 0.46
in the 2-4 and 4-10 keV folded light curves, respectively,  the HR
increases from 0.92 to 1.01
We conclude that a periodic
  signal at a period of 24.53 d is present both in the 1.3-12.1 keV ASM and
  2-10 keV MAXI light curve; furthermore, the 24.53 d folded light curves
  highlights the presence of a periodic
  dip.

\subsection{EXOSAT}   

\begin{figure}
\includegraphics[angle=-90,width=\columnwidth]{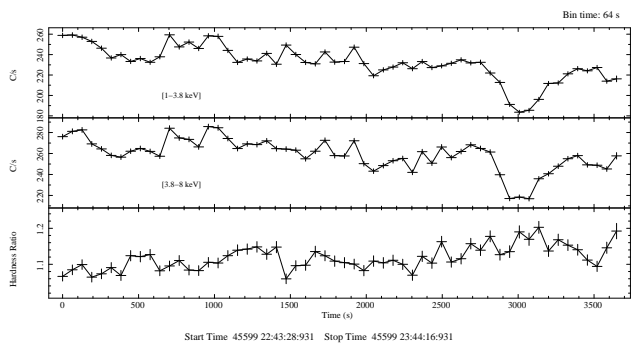}
\caption{EXOSAT/ME observation of GX 13+1 between 45,599.95 and
  45,599.99 MJD.  Light  curves in the 1-3.8 keV (top panel),
   in the 3.8-8 keV (middle panel) energy band, and the
    corresponding hardness ratio (bottom panel) are shown. The bin
    time is 64 s.}
\label{Fig3}
\end{figure}

\begin{figure}
\includegraphics[angle=-90,width=\columnwidth]{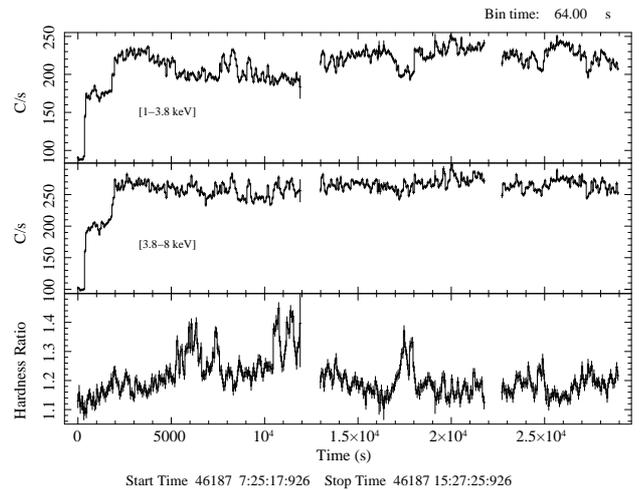}
\caption{EXOSAT/ME observation of GX 13+1 between  46,187.31 and
  46,187.64 MJD.  Light curves in the 1-3.8 keV (top panel),  in the
  3.8-8 keV (middle panel) energy band, and the corresponding hardness
  ratio (bottom panel) are shown}. The bin time is 64 s.
\label{Fig4}
\end{figure}

Encouraged by these results, we looked for historical observations of
GX 13+1, in order to find periodic dips in pointed observations,
starting with EXOSAT.  We used the Medium energy (ME) light curves in
the NASA archive\footnote{The standard filtering of the ME light
  curves is shown at
\url{heasarc.gsfc.nasa.gov/W3Browse/exosat/me.html}}
The ME instrument on board EXOSAT observed GX 13+1 on 1983 Sep. 22
(seq. number 192), 1985 Apr. 1 (seq. number 1488) and 1985 May 2
(seq. number 1549).
We analysed the corresponding 1-3.8 and 3.8-8 keV light curves and the
corresponding HR; the light curves
at the two different energy bands and the HRs are shown in
Fig. \ref{Fig3} and \ref{Fig4}.

We note that the light curves corresponding to the observation taken
on 1983 Sep. 22 (Fig. \ref{Fig3}) show a decrease of count rate at
around 3,000 s from the start time in both the energy bands and the
corresponding HR changes slightly going from 1.1 to 1.2.  The light
curves corresponding to the observation taken on 1985 May 2 (see
Fig. \ref{Fig4}) show a long-term modulation of the hardness ratio
that it is difficult to ascribe to the presence of a dip. During the
first 15 ks from the beginning of the observation the HR increases from
1.1  to 1.4; at 17.5 ks from the start time  it
rapidly increases from 1.1 to 1.3, in the last part of the observation
it is quite constant at 1.15. Although it is evident a change of the
HR in the EXOSAT observation taken in 1985 we are not able to ascribe
this behaviour to a dip. For this reason in our analysis we will not
use the EXOSAT observations.
\subsection{XMM/RGS}   
\label{rgs}
\begin{figure}
\includegraphics[angle=-90,width=\columnwidth]{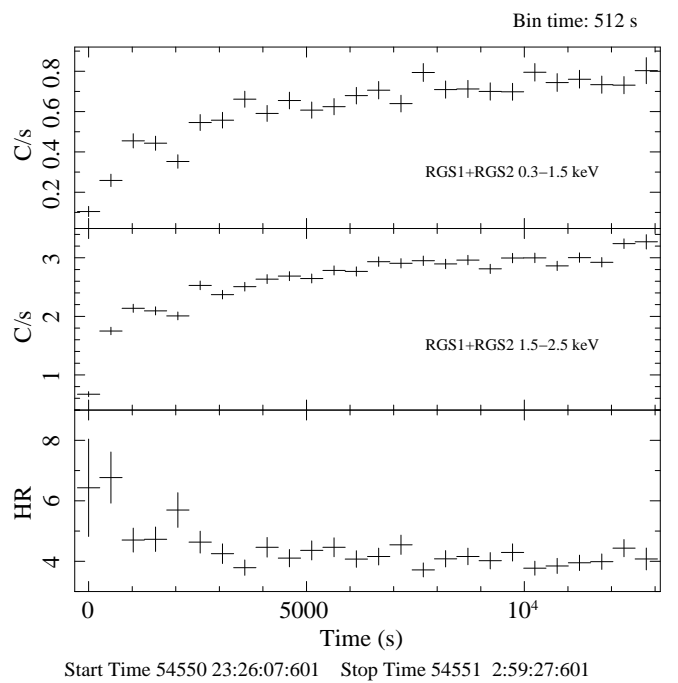}
\caption{RGS  light  curves  during  the  observation  ID  0505480501,
  between 54,550.97  and 54,551.12 MJD.  Upper panel:  0.3-1.5 keV RGS
  light curve. Middle Panel: 1.5-2.5 keV RGS light curve. Lower panel:
  the corresponding HR.  The bin time is 512 s.}
\label{Fig5}
\end{figure}
\begin{figure}
\includegraphics[angle=0,width=\columnwidth]{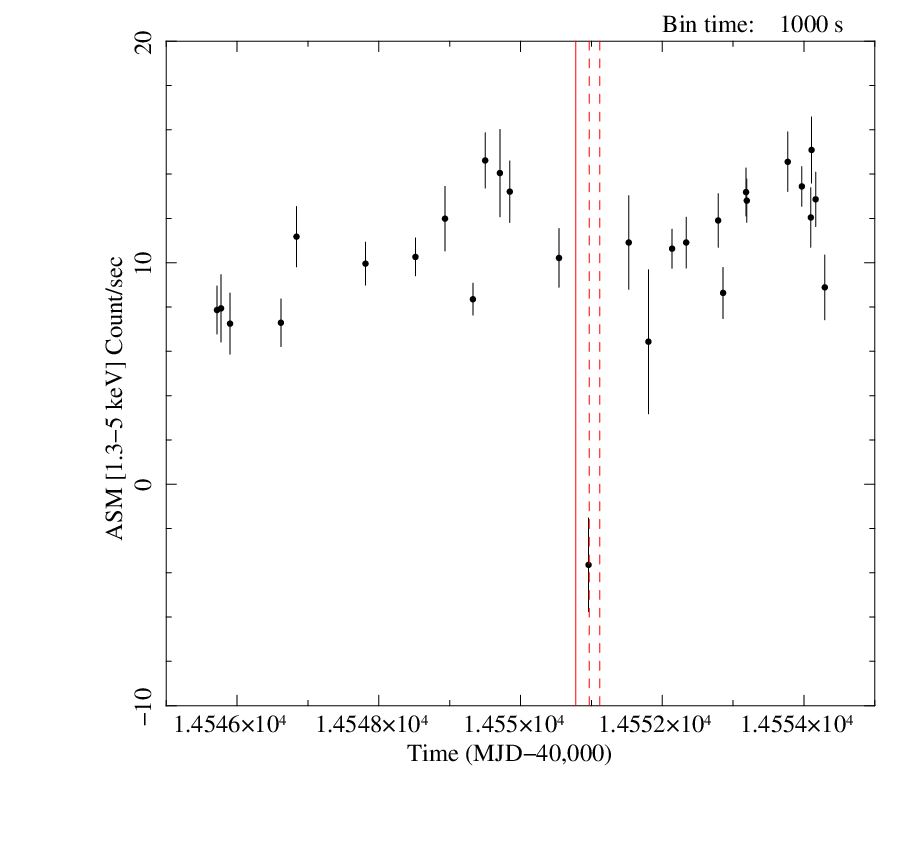}
\caption{RXTE/ASM light curve  of GX 13+1in the 1.3-5  keV energy band
  between 54,545 and  54,555 MJD.  The bin time is  1000 s. The dotted
  vertical  lines  indicate  the  start  and stop  times  of  the  RGS
  observation.   The solid  vertical line  indicates the  expected dip
  arrival  time  (i.e.   54,550.78  MJD)  adopting  the  ephemeris  of
  eq. \ref{unofit_eq} (see the text).}
\label{Fig5b}
\end{figure}
XMM-Newton observed GX 13+1 nine times; a complete analysis of the RGS
and Epic-pn light curves and spectra of the source has been recently
reported by \cite{Trigo_2012}. The most significant detection of a  
dip is during the observation ID 0505480501 taken on 2008 March
25. However, the dip was observed only partially by the RGS because it
occurred at the beginning of the observation while the EPIC-pn was
still off. The start and stop times of the RGS observation
are 54,550.97 and 54,551.12 MJD, respectively. 
We show the background-subtracted light curves of
the summed 1st-order RGS1 and RGS2  data in the 0.3-1.5 keV
(Fig. \ref{Fig5}, upper panel) and 1.5-2.5 keV (Fig. \ref{Fig5},
middle panel) energy band. The corresponding HR is shown in the lower
panel of Fig. \ref{Fig5}.  The HR is the hardest (about 7) during the
first 1,000 s of the observation, then drops to a value of 4,
suggesting that the decrease of the RGS count rate at the beginning of
the observation is associated with a dipping phenomenon.

We do not have a complete coverage of the dip and 
are not able to give a precise estimate of
the dip arrival time. To get more
information on this dip, we looked for it in the 1.3-5 keV RXTE/ASM light
curve.  We show the RXTE/ASM light curve between 54,545 and 54,555 MJD
in Fig.  \ref{Fig5b}; the start and stop times of the RGS observation
are indicated with the red-dashed vertical lines.
The ASM light curve of GX 13+1 shows that the RGSs observed only the
end of the dip. We will not take into account this dip for our
analysis, because we were not able to estimate exactly its arrival
time. However, we will comment a-posteriori our results comparing them
with the ASM light curve shown in Fig. \ref{Fig5b}.

\section{Ephemeris of GX 13+1}
 \begin{figure}
\begin{centering}
\includegraphics[angle=0,height=5.5cm,width=\columnwidth]{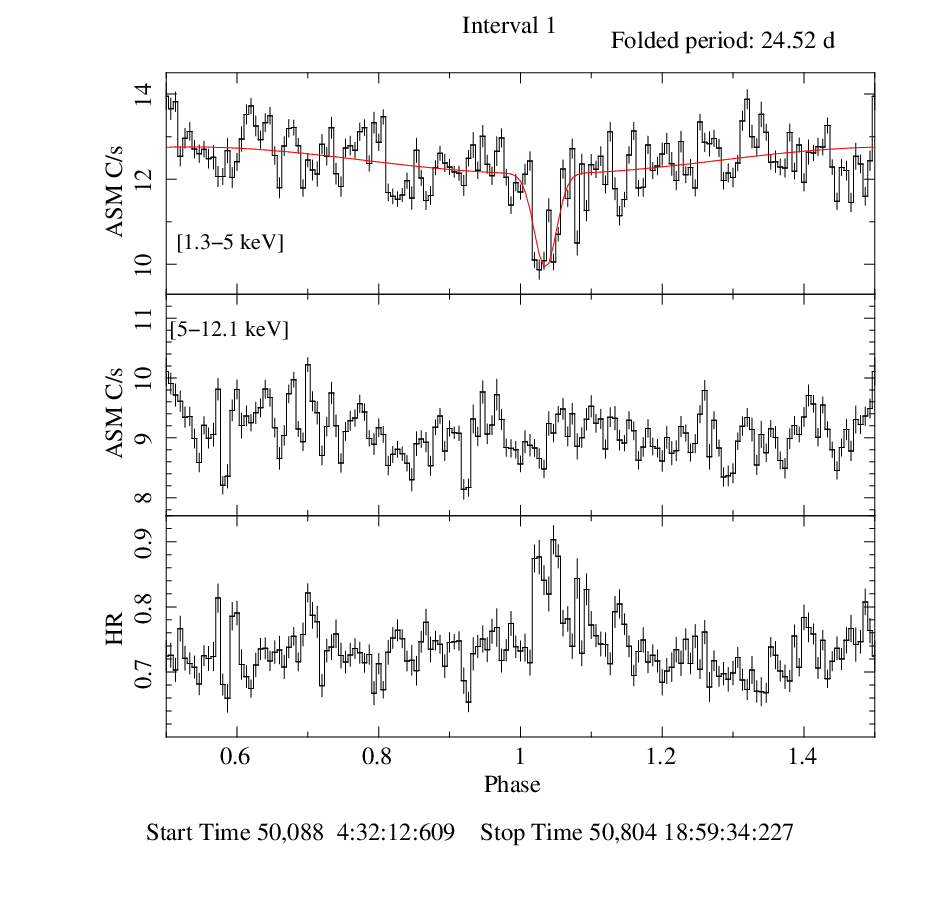}
\includegraphics[angle=0,height=5.5cm,width=\columnwidth]{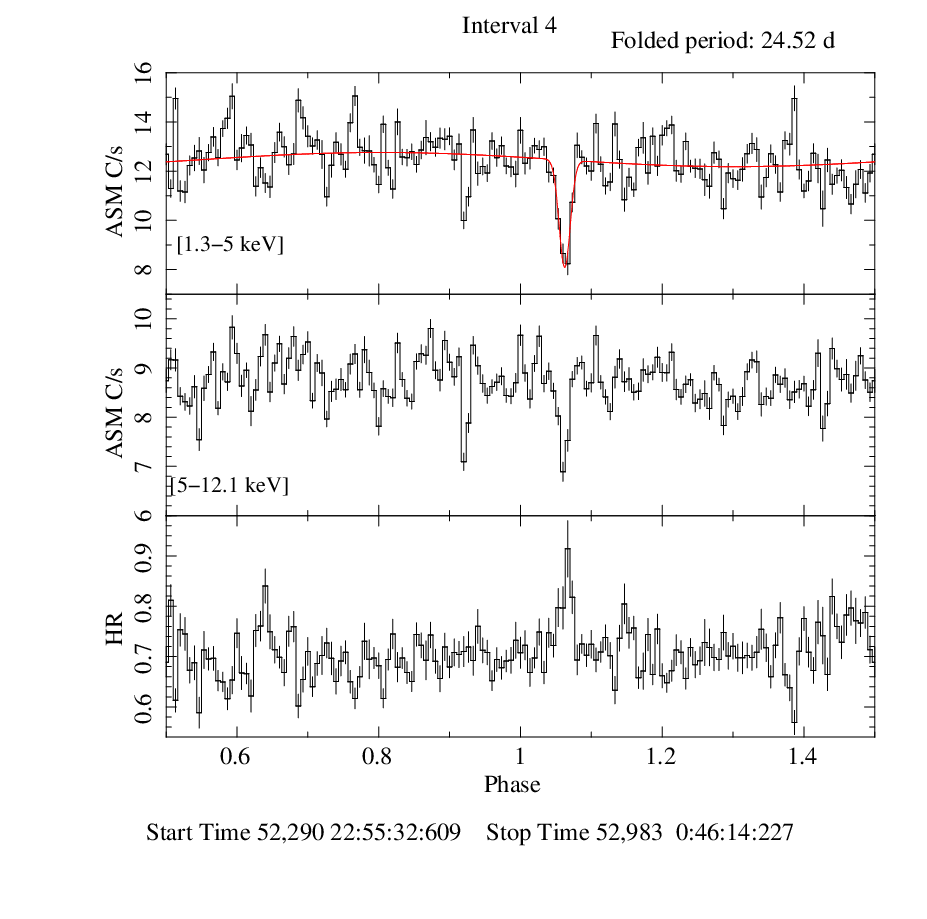}
\includegraphics[angle=0,height=5.5cm,width=\columnwidth]{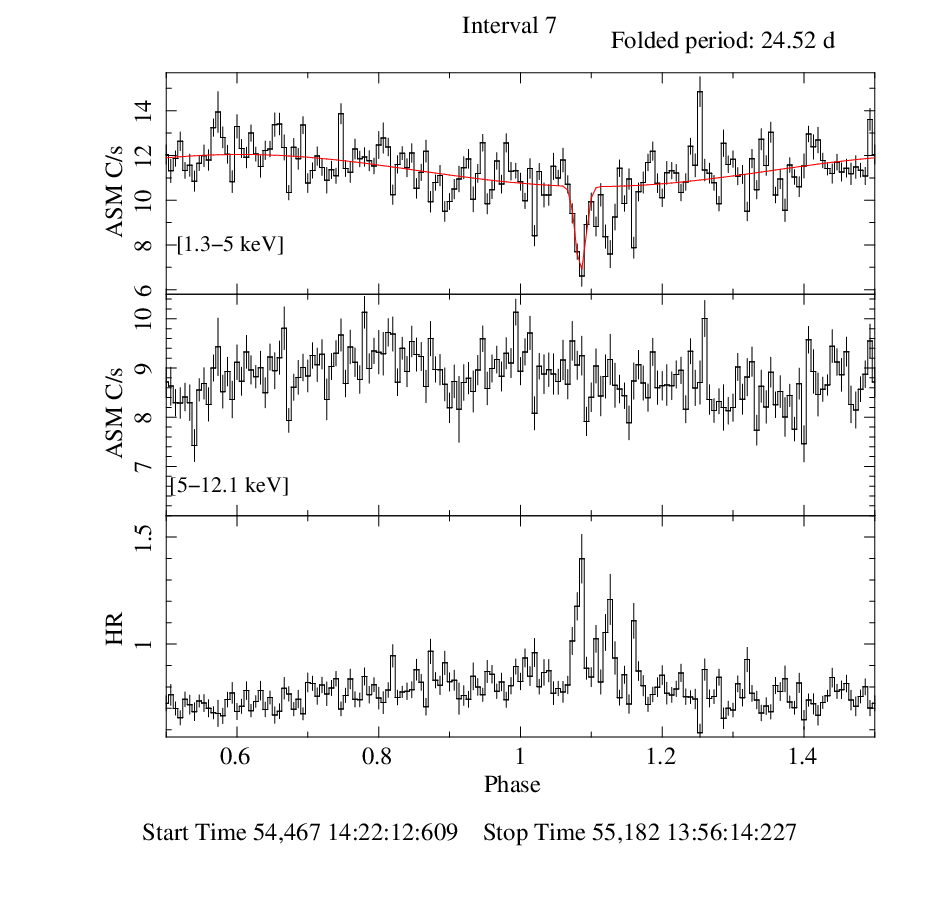}
\caption{ASM folded light curves of GX 13+1 in the 1.3-5 keV (top
  panels) and 5-12.1 keV energy band (middle panels) covering about
  714 days.  The corresponding hardness ratios are plotted in the
  bottom panels.  The folding period is 24.52 d and 150 phase-bins per
  period were adopted. The red solid line is the best-fit curve
  composed of a constant plus a sinusoidal function with period kept
  fixed to 24.52 d and a Gaussian profile fitting the  dip.}
\label{Fig6int}
\end{centering}
\end{figure}
\begin{figure}
\begin{centering}
\includegraphics[height=5cm,angle=0,width=\columnwidth]{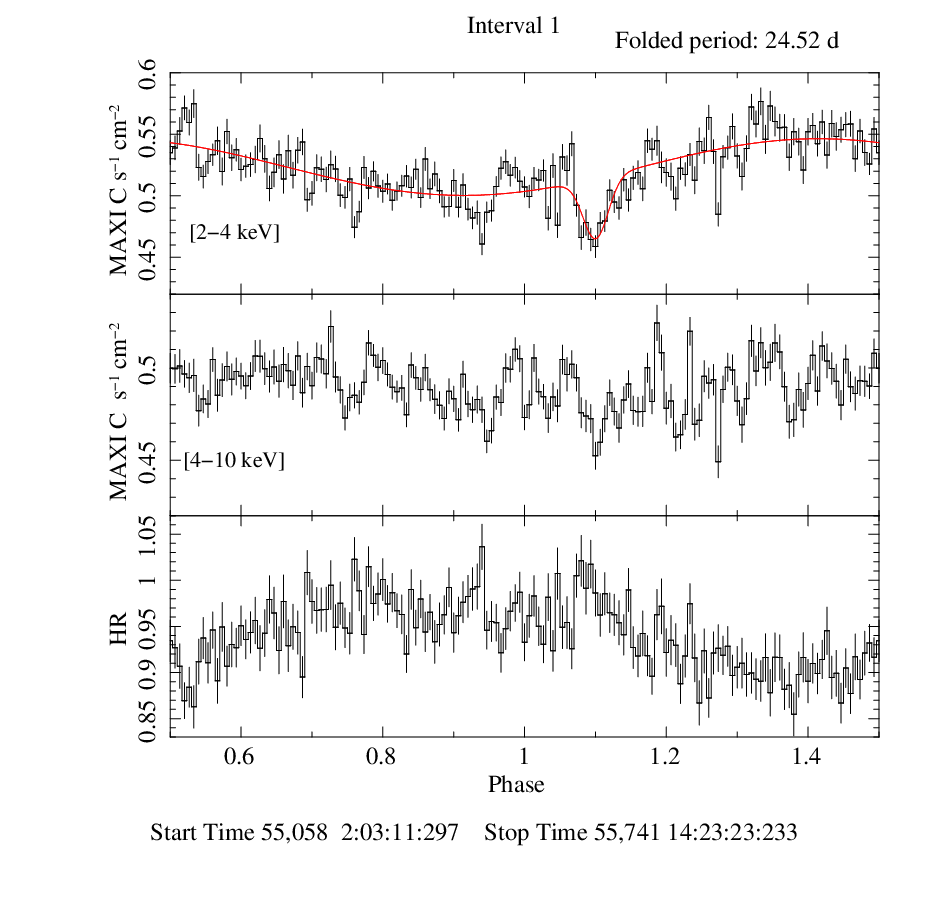}
\includegraphics[height=5cm,angle=0,width=\columnwidth]{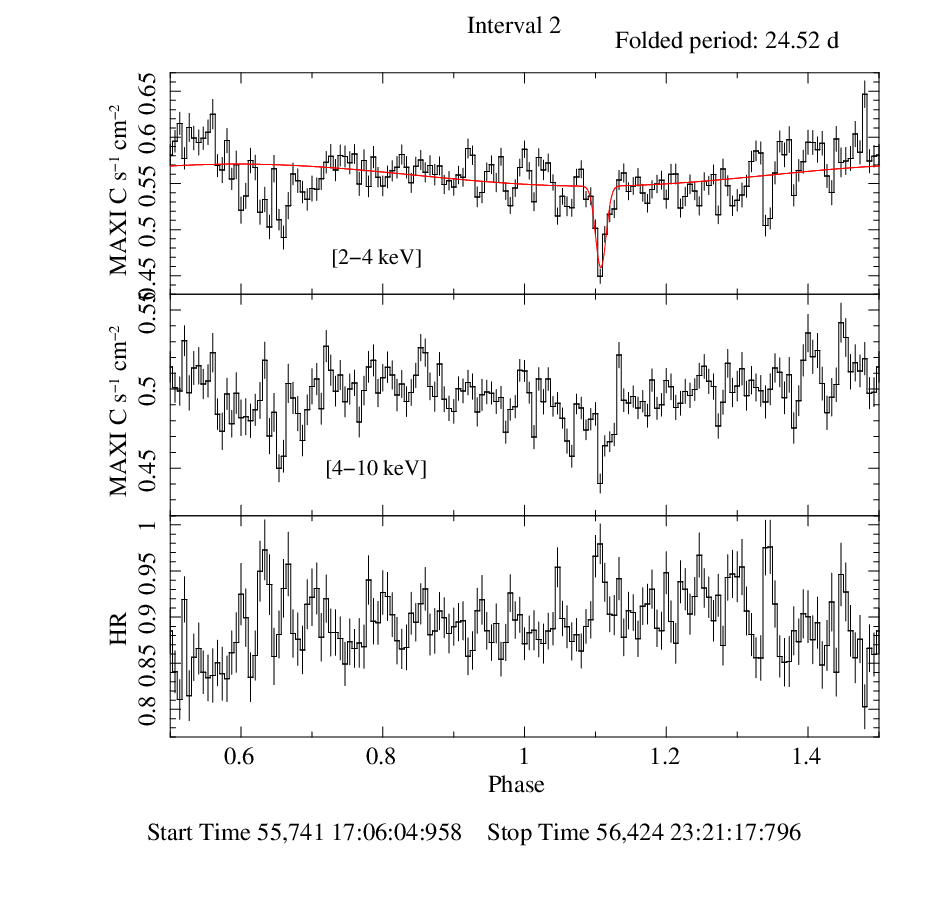}
\caption{MAXI folded light curves of GX 13+1 in the 2-4 keV (top panels) and 
4-10 keV energy band (middle panels)  
covering about 680 days. The folding period is
  24.52 d and 150 phase-bins per period were adopted. The red solid line 
is defined in the text and Fig. \ref{Fig6int}. }
\label{Fig6intb}
\end{centering}
\end{figure}

To estimate with the best accuracy the period of GX 13+1 we used
a timing technique (see e.g. \cite{Iaria_11} and references
  therein, for an example of the application of timing techniques to
  improve orbital ephemeris). In order to obtain several dip arrival
times we divided the 1.3-5 keV and 5-12.1 keV ASM light curves in
eight  intervals.  The corresponding temporal windows of each of
the eight light curves cover about 29 cycles ($\sim 714$ days). 
  For each interval, the soft and hard light curves were folded
adopting an arbitrary folding period of 24.52 d (close to 24.53 d
observed in the ASM and MAXI light curves), an arbitrary reference
time of 50,086 MJD, and, finally, 150 phase-bins per period.  The
presence of a periodic dip in the folded light curves was 
  significantly detected only in intervals 1, 4, and 7
(Fig. \ref{Fig6int}), the start and stop times of these intervals are
shown in Tab. \ref{Tab1}.  To estimate the dip arrival times we fitted
the 1.3-5 keV folded light curves with a model consisting of a
constant plus a sinusoidal function, with the period kept fixed to 1
(that is 24.52 d), adding a Gaussian component with a negative
normalisation to fit the dip shape and assuming the centroid of
the Gaussian component as the dip arrival times.  We took care to fit
only the dip in the 1.3-5 keV folded light curve showing an evident
increase of the HR $[5-12.1 {\rm keV}]/[1.3-5 {\rm keV}]$.
\begin{table*}
  \caption{Journal of the X-ray dip arrival times of GX 13+1 \label{Tab1} }
\begin{center}
\begin{tabular}{c c c l  c l  c}          
\hline\hline 
Interval &
Tstart &
Tstop &
Dip Time  &               
Cycle  &
Delay &
Satellite \\
  
 &
 (MJD;TDB) &               
 (MJD;TDB) &               
 (MJD;TDB) &               
  &
(days)&
 \\

\hline                        

 1 &
 50,088.19 &
 50,804.79 & 
  50,454.66(8)&
 15&
 0.86(8)&
RXTE/ASM\\ 

4 &
 52,290.96 &
 52,983.03  & 
 52,637.61(6)&
 104&
 1.53(6)&
RXTE/ASM\\ 

7 &
  54,467.60  &
 55,182.58   & 
 54,844.97(6)&
 194&
 2.09(6)&
RXTE/ASM\\

 &&&
 55,409.22935(11)&
217&
 2.38935(11)&
Chandra/HETGS\\ 

1 &
 55,058.09 &
  55,741.47 & 
 55,409.30(15)&
 217&
 2.45(15)&
MAXI\\ 

2 &
 55,741.59 &
 56,424.97   & 
 56,096.04(6)&
 245&
 2.67(6)&
MAXI\\

\hline                                             
\end{tabular}
\end{center}

{\small \sc Note} \footnotesize--- Epoch of reference $ 50,086$ MJD, 
orbital period $24.52$ days.
\end{table*}
\begin{figure*}
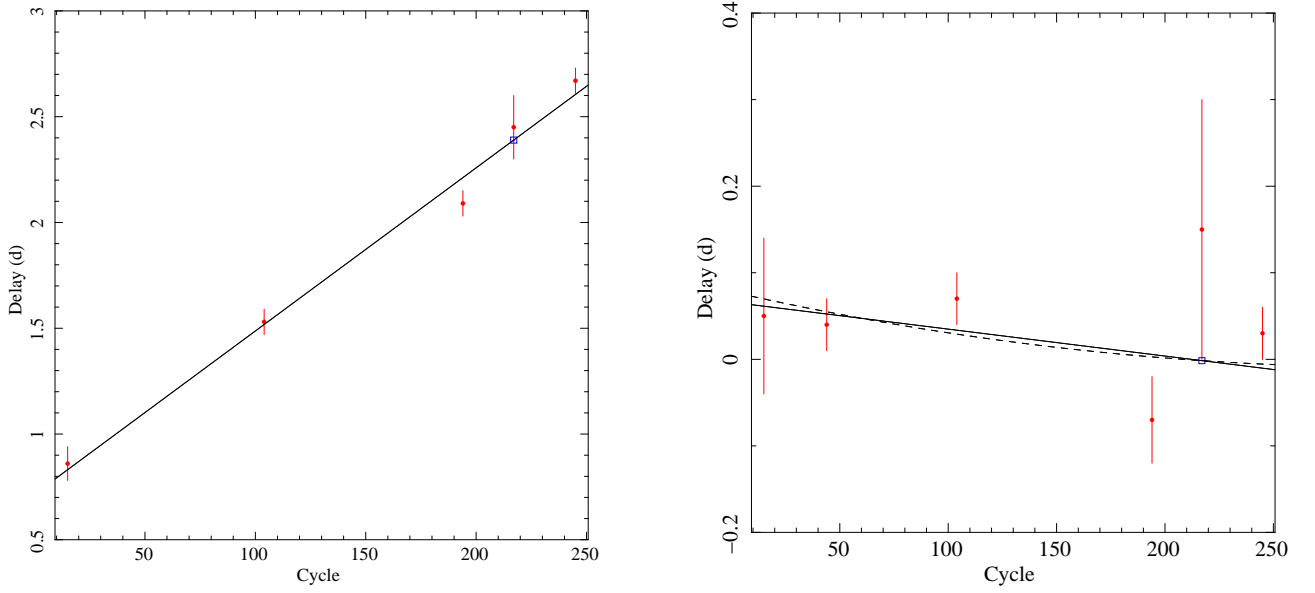

\includegraphics[angle=0,width=\columnwidth]{FINAL_DELAY_PLOT_PAPER_10_JUL_ERRORX_fit.ps}
\includegraphics[angle=0,width=\columnwidth]{FINAL_DELAY_PLOT_PAPER_10_JUL_245277_triangolochandra.ps}
\caption{Delays in units of days of the dip arrival times plotted versus the corresponding cycle number; the Chandra point is indicated with 
a blue open square. Left panel: trial period of 24.52 d and epoch  
   50,086  MJD; the solid line  represents the
  linear ephemeris shown in eq. \ref{zerofit_eq}.
 Right panel: trial period of 24.5277 d and  epoch 50,086.72
  MJD. The solid line and the dashed curve indicate 
 the linear ephemeris shown in Eq. \ref{unofit_eq}
  and the quadratic ephemeris of Eq. \ref{secondofit_eq}, respectively.}
\label{primofit}
\end{figure*}
\begin{table*}
  \caption{Journal of the refined X-ray dip arrival times of GX 13+1 \label{Tab2} }
\begin{center}
\begin{tabular}{c c c l  c l  c}          
\hline\hline 
Interval &
Tstart &
Tstop &
Dip Time  &               
Cycle  &
Delay &
Satellite \\
  
 &
 (MJD;TDB) &               
 (MJD;TDB) &               
 (MJD;TDB) &               
  &
(days)&
 \\

\hline                        

 1 &
  50,088.19 &
 50,804.79 & 
  50,454.69(9)&
 15&
0.05(9)&
RXTE/ASM\\

2 &
 50,816.23  &
  51,506.13 & 
  51,165.98(3)&
 44&
 0.04(3)&
RXTE/ASM\\ 

4 &
 52,290.96  &
 52,983.03  & 
 52,637.67(3)&
 104&
 0.07(3)&
RXTE/ASM\\ 

7 &
 54,467.60  &
 55,182.58   & 
 54,845.16(5)&
 194&
 -0.07(5)&
RXTE/ASM\\

 &&&
 55,409.22935(11)&
217&
 -0.00157(11)&
Chandra/HETGS\\

1 &
 55,058.09 &
  55,741.47 & 
 55,409.38(15)&
 217&
 0.15(15)&
MAXI\\ 

2 &
55,741.59 &
 56,424.97   & 
 56,096.03(3)&
 245&
 0.03(3)&
MAXI\\

\hline                                             
\end{tabular}
\end{center}

{\small \sc Note} \footnotesize--- Epoch of reference $ 50,086.72$ MJD, 
orbital period $ 24.5277$ days.
\end{table*}

Adopting the same procedure we divided the 2-4 keV MAXI light curve
into two intervals, the corresponding temporal windows of each of the two
MAXI light curves cover about 28 cycles ($\sim 680$ days). We show the
MAXI folded light curves in Fig. \ref{Fig6intb} and  report the journal of the X-ray dip arrival times in Table \ref{Tab1}.

This analysis allows us to obtain five dip arrival times from the ASM
and MAXI light curves. Note that, in order to increase the statistics in the ASM
and MAXI light curves,  we selected large temporal windows,
covering  29 and 28 cycles in the RXTE/ASM and MAXI
light curves, respectively.  
We adopted  as Epoch of reference the arbitrary value of 50,086 
MJD and calculated the delays of the dip arrival times with respect to
those predicted by 
the trial folding period of 24.52 d.
We assigned to each temporal window the  cycle number 
corresponding  to the half time of the window.
We, initially, fitted the delays  vs orbital cycle with a linear function $f$ 
and took into account 
 only the errors associated to the dip arrival times (hereafter
$\Delta y$)   

   obtaining 
  a first estimation of the corrections to the epoch of reference
  and orbital period.  Since each temporal window contains a large number of
  cycles we took into account the uncertainty associated to this large
  number combining $\Delta y$ with the error associated to the cycles
  (hereafter $\Delta x$) using the relation $(\Delta _{tot})^2 =
  (\Delta y)^2 + (\partial f/\partial x \;\Delta x)^2$ and fitted
  again the delays vs cycles. $\Delta _{tot}$ are the errors
associated to the dip arrival times shown in Tab. \ref{Tab1}.

We obtained six dip arrival times one from Chandra, two from the
folded MAXI light curves, and, finally, three from the folded RXTE/ASM
light curves.  We corrected the orbital period fitting the delays with
a linear function.  The linear ephemeris is
\begin{equation}
\label{zerofit_eq}
T_{dip} = 50,086.72(7) {\; \rm MJD}+  24.5277(3) N, 
\end{equation}
where $N$ is the number of cycles, $ 50,086.72(7)$ is the new Epoch of
reference, and the revised orbital period is $P= 24.5277(3)$ d. The
$\chi^2({\rm {d.o.f.}})$ is 5.8(4) and the associated errors are at 
68\% confidence level. Here and in the following the uncertainties in
the parameters have been scaled by a factor $\sqrt{\chi^2_{red}}$ to
take into account a $\chi^2_{red}$ of the best-fit model larger than
1. We show  the delays versus the orbital cycles, together with 
the best-fit linear ephemeris, in Fig.
\ref{primofit} (left panel).

We also determined the linear ephemeris excluding the Chandra point to
avoid the possibility that our result is driven by it. In this case we
found that the new epoch of reference is $ 50,086.73(9)$ and the
revised orbital period is $P= 24.5276(5)$ d. This result is compatible,
within 1 $\sigma$,
with that shown in Eq. \ref{zerofit_eq} and suggests that the Chandra
point does not drive the fit and that the dip observed in the Chandra
light curve  is  periodic. Furthermore, we started with a trial
period of 24.52 d obtaining, finally, a period of 24.5277 d that is
very similar to the value of 24.53 d obtained independently searching
for periodicities in the ASM and MAXI light curves (see Figs. \ref{win_2}
and \ref{maxi_efsearch}).

Then, we applied again the whole procedure to the ASM and MAXI light curves
adopting as folding period   24.5277 d and as epoch of reference 50,086.72 MJD
to refine our results. The selected intervals for the two light curves 
are the same described  above. Using these two accurate values we were 
able to detect a periodic dip also in the interval 2 of the ASM light curves.
The journal of the seven dips is shown in Tab. \ref{Tab2}.

\begin{figure*}
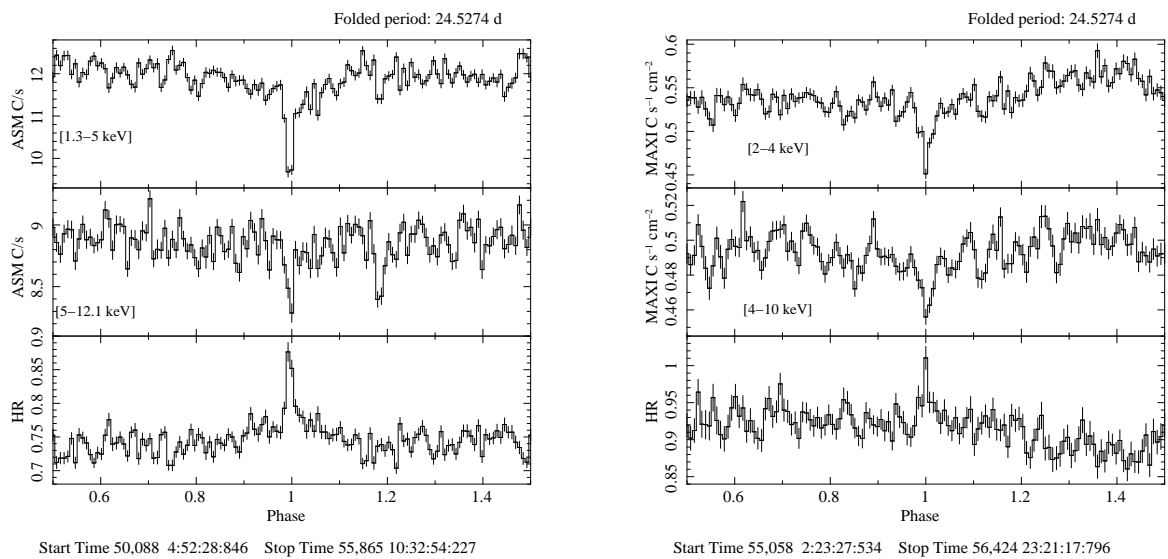

\includegraphics[height=8cm,angle=0]{final_fold_ASM_128binned_245274.ps}
\includegraphics[height=8cm,angle=0]{folded_soft_hard_24d5274_binned128.ps}
\caption{RXTE/ASM (left panel) and MAXI (right panel) 
folded light curves and the corresponding hardness ratios 
using the ephemeris of  Eq. \ref{zerofit_eq}.  The RXTE and MAXI 
 light curves are folded 
using 128 phase-bins per period.}
\label{Fig7}
\end{figure*}

We corrected the orbital period fitting the seven delays with a linear
function. The linear fit of the seven points is shown in
Fig. \ref{primofit} (right panel).  The refined linear ephemeris is
\begin{equation}
\label{unofit_eq}
T_{dip} = 50,086.79(3) {\; \rm MJD}+  24.5274(2) N.
\end{equation}
The $\chi^2({\rm {d.o.f.}})$ is 6.76(5) and the associated errors are
at  68\% confidence level.  We folded the 1.3-5 keV and 5-12.1 keV
ASM light curves adopting the epoch of references and the period of
Eq. \ref{unofit_eq} (Fig. \ref{Fig7}; left panels), both the light
curves cover $\sim 236$ cycles.  The periodic dip is evident in both
the ASM energy bands and the HR shows at zero phase  a more pronounced
increasing. To quantify the improvement of the orbital period after the 
correction achieved using the timing technique we defined   the fraction
$f=(HR_{max}-HR_{ave})/(HR_{max}+HR_{ave})$, where $HR_{ave}$ is the
average value of HR between phase 0.1 and 0.9 and $HR_{max}$ is the HR
at phase 0.  We obtain that $f$ is 6.1\% and 8.6\% in the HR obtained
folding the ASM light curves with a period of 24.53  and 24.5274 d,
respectively. 

We folded the 2-4 keV and 4-10 keV
MAXI light curves adopting the epoch of references and the period of
Eq. \ref{unofit_eq} (Fig. \ref{Fig7}, right panels), both the light
curves covers $\sim 56$ cycles.
The value of $f$ is 4.5\% and 5\% in the HR obtained
folding the MAXI light curves with a period of 24.53  and 24.5274 d,
respectively.  The largest
increase of the HR associated with the ASM light curves is mainly due
to the wider spanning time that ASM covers ($\sim 16$ yr) with respect
to MAXI ($\sim 4$ yr).

Finally, we added a quadratic term to the ephemeris to take into account 
a possible period derivative. In this case, we obtained  the following 
quadratic ephemeris
\begin{equation}
\label{secondofit_eq}
T_{dip} =  50,086.80(7) \;{\rm MJD}+  24.5271(13) N  
 +9(46)\times 10^{-7} N^2, 
\end{equation}
with a $\chi^2({\rm {d.o.f.}})$ of  6.7(4).  Adding the
quadratic term we do not obtain a significant improvement of the fit.
The corresponding period derivative is $ 8(37) \times 10^{-8}$
s/s with the associated error  at 68\% confidence level. The
best-fit curve associated to the quadratic ephemeris is shown in Fig.
\ref{primofit} (right panel) with a dashed curve.
\begin{table*}
  \caption{Journal of the XMM observations of GX 13+1 \label{Tab3} }
\begin{center}
\begin{tabular}{l c  c c c c}          
\hline\hline 
Obs. & Observation   &               
\multicolumn{2}{c} 
{Start Time  (UTC)} 
 &
Orbital Phase & Dipping State\\

Num. &
ID &               
year month day  &
hr:min &
 \\

\hline                        

 1 &
0122340101&
2000 March 30&
 14:10&
0.06&
deep
\\ 

2&
0122340901&
2000 April 01&
4:29& 
0.13& shallow\\

3&
0122341001&
2000 April 01&
8:51&
0.14& deep
\\ 

4&
0505480101&
2008 March 09&
15:48&
0.34& shallow\\ 

5&
0505480701&
2008 March 11&
19:00&
0.43&--\\

6&
0505480201&
2008 March 11&
23:10&
0.44& persistent\\ 


7&
0505480301&
2008 March 22&
02:20&
0.85& shallow\\ 

8&
0505480501&
2008 March 25&
23:01&
0.007 & deep\\ 

9&
0505480401&
2008 September 5&
21:49&
0.69 & deep\\

\hline                                             
\end{tabular}
\end{center}

{\small \sc Note} \footnotesize--- 
The phase is associated 
to the start time of the observation and it is calculated using the 
ephemeris of Eq. \ref{unofit_eq}. The dipping state shown in the sixth column
is discussed by   \cite{Trigo_2012}. For the observation 5 only RGS and the optical monitor (OM) were available.
\end{table*}

Using the linear ephemeris of Eq. \ref{unofit_eq} we determined the
expected dip arrival time nearest the RGS observation discussed in
section 2.4. The corresponding cycle number is 182 and the predicted
arrival time is 54,550.78 MJD.  In Fig. \ref{Fig5b} we show the
RXTE/ASM light curve between 54,545 and 54,555 MJD. We indicate the
start and stop times (54,550.97 and 54,551.12 MJD, respectively) of
the RGS observation with the red-dashed vertical lines; the predicted
dip arrival time using the ephemeris of Eq. \ref{unofit_eq} is plotted
with a red-solid vertical line.  The separation in units of
orbital phase of the expected dip arrival time from the RGS start time
is only 0.19 d.  We conclude that our predicted arrival time is
compatible with the ASM light curve and that the RGS instruments
observed only the final part of a periodic dip.

 \cite{Trigo_2012} suggested the presence of possible dipping states
in RXTE/PCA, EXOSAT, ans XMM observations.
We,  therefore, searched for possible periodic dips in the RXTE/PCA
archival data.  We analysed the barycentred 2-4 and 4-9 keV light
  curves obtained from the standard 2 products finding that there are
no observations covering the phase zero (i.e. the phase at which we
expect the dip), except for observation P95338-01 that unfortunately
has a hole in the light curve at the expected dip arrival time (see
sec. 2.1).
  The observation P95338-01 spans 12 days covering almost one half of
  the orbital period.  The hardness ratio ranges between 2 and 2.6,
  reaching the maximum value at phases 0.95 and close to 1.

\cite{Trigo_2012} suggested a possible dipping state of GX 13+1
observed with the RXTE/PCA in the observations P30050-01 and
P30051-01. However, the authors did not perform a systematic search
for dips and their suggestion is based on the bi-modal behaviour of
the count rate with the hardness ratio reported by \cite{Schnerr_03}.  We
reanalysed the RXTE/PCA observations P30050-01 and P30051-01 which
span a time interval from 50,950 to 51,096 MJD (i.e. 5.94
orbital cycles for a period of 25.5274 d). We analysed the
barycentred 2-4 and 4-9 keV light curves and extracted the corresponding HR.
 We found  that 
the HR has an erratic behaviour along the
orbital phase varying between 2.5 and 3.5 and reaching the maximum
value of 3.8 at phase 0.38. The observations do not cover
the phase at which the periodic dip is expected; the pointed
observations closest to the zero-phase are P30051-01-06-00 and
P30051-01-10-00 (phase at the start time 0.98 and 0.97, respectively).
Finally, we also analysed  the RXTE/PCA observations P40023-03
and P40022-01 which span $\sim 171$ days  (i.e. $\sim 7$ orbital
periods). The HR ranges between 1.8 and 2.2 and has a peak
at phase 0.19 reaching the value of 2.7.

\cite{Trigo_2012} also  suggested the  presence of a  possible dipping
state of GX 13+1 in the  EXOSAT observation taken on 1983 Sep. 22 (see
sec. 2.1); using  the linear ephemeris of eq.  \ref{unofit_eq} we find
that  the start  time of  that observation  corresponds at  phase 0.07
meaning that the dip arrival time  is predicted  at $\sim 144$ ks
before the start  time), that is quite far  from the expected periodic
dip  arrival  time.   The  EXOSAT  observation taken  on  1985  May  2
corresponds to an orbital-phase interval 0.015-0.029, that is $\sim$33
ks after the expected periodic dip.

The  presence of  dipping  states was  also  suggested in  XMM/Epic-pn
observations. For  sake of completeness  we report in  Tab. \ref{Tab3}
the list  of the XMM  observations of GX  13+1 using the  same indices
adopted  by \cite{Trigo_2012}  in their  paper.  We  added  as further
information  the  orbital  phase  associated   to  each  of  the
XMM/Epic-pn start time adopting  the ephemeris of eq. \ref{unofit_eq}.
In this  case we converted in phase  the start time in  UTC instead of
TDB; however, in our case, the effects of the barycentre correction is
irrelevant because of the large period of 24.5274 d.  We note that the
dipping states  of GX 13+1  suggested by \cite{Trigo_2012}  are almost
independent on the orbital phase.

Finally, Chandra observed GX 13+1 nine times; we report in Tab. \ref{Tab4}
the list of the observations and the orbital phase at the start time of
each Chandra observation. 
\begin{table}
  \caption{Journal of the Chandra observations of GX 13+1 \label{Tab4} }
\begin{center}
\begin{tabular}{l  l c c}          
\hline\hline 
Observation   &               
\multicolumn{2}{c} 
{Start Time  (UTC)} 
 &
Orbital Phase \\

ID &               
year month day  &
hr:min:sec
 \\

\hline

950&
2000 August 7&
 00:00:05&
0.34
\\

2708&
2002 October 8&
11:12:56& 
0.65\\

6093&
2005 February 8&
19:50:44&
0.48
\\

11815&
2010 July 24&
05:46:27&
0.67\\

11816&
2010 July 30&
14:47:25&
0.93\\

11814&
2010 August 1&
00:31:31&
0.99\\

11817&
2010 August 3&
10:12:10&
0.09 \\

11818&
2010 August 5&
14:09:39&
0.18 \\

13197&
2011 February 17&
17:57:04&
0.18 \\

\hline                                             
\end{tabular}
\end{center}

{\small \sc Note} \footnotesize---  
The phase is associated 
 to the start time of the observation and it is calculated using the 
ephemeris of Eq. \ref{unofit_eq}. The observation id. 11814 is discussed in 
sec. 2.1. 
\end{table}

\section{Discussion and Conclusion}
In this paper we have analysed archival data of GX 13+1 from Chandra, XMM,
RXTE, EXOSAT, and MAXI, with the aim to look for periodic dips in this source.
Using a Chandra/HETG observation
we improved the X-ray position of GX 13+1, that is now compatible with
the positions of the infrared and radio counterparts of GX 13+1. Using
the new coordinates we barycentred the ASM and MAXI light curve of GX
13+1.  We performed a PSD of the 1.3-12.1 keV ASM light curve
detecting a periodic signal  at a period between 22.78 and 26.92
d. To refine the periodic signal we used the ftool {\it efsearch} and
looked at the period-window between 22.78 and 26.92 d in the 1.3-12.1
keV ASM and 2-10 keV MAXI light curves.  We detected a significant
common periodic signal in the two light curves at 24.53 d.  The 24.53
d folded ASM and MAXI light curves in two different energy bands
(1.3-5 and 5-12.1 keV for ASM and 2-4 and 4-10 keV for MAXI) and the
corresponding HRs clearly show that the periodic signal is associated to
a periodic dip present in the light curves.
  
To improve the value of 24.53 d and to obtain the orbital ephemeris of
GX 13+1 we used the timing technique dividing the ASM and MAXI light
curves in eight and two time intervals, 
respectively. Finally we obtained four dip
arrival times from the ASM light curves and two dips arrival times 
from the MAXI light curve. We showed that the dip observed in a recent
Chandra observation is well fitted using the ephemeris obtained by
our dip arrival times. The linear ephemeris gave a refined period of
24.5274(2) d that we interpret as the orbital period of the system.

So far, the absence of periodic signatures in the light curve of GX
13+1 has made difficult an accurate estimation of its orbital period.
Before this work, the most accurate estimation of the orbital period
of the system was given by \cite{Corbet_2010}, who looked for
periodicities in the RXTE/ASM light curve spanning 14 years, finding a
period of $24.27 \pm 0.03$ days that is not compatible with our
result; however, the associated error to the period reported by
\cite{Corbet_2010} is a formal statistical error and it could be
larger. The detection of periodic dips allows us to determine accurate
X-ray ephemeris of GX 13+1 and improve the accuracy of the period by a
factor one hundred.


From the analysis of the delays associated with the dip arrival times
we obtained the linear and quadratic ephemerides of GX 13+1. The
  goodness of our orbital solution is evident when we fold the whole
  RXTE/ASM and MAXI light curves (Fig. \ref{Fig7}) after correcting
  the event arrival times for the new orbital solution, since in this
  case both the light curves show a clear dip at zero phase with a
  corresponding increase of the HRs.  If the proposed period of
  24.5274 d, obtained by studying the periodic dips in the light curve
  of GX 13+1, will be confirmed as the orbital period of the system,
  then GX 13+1 has the longest known orbital period for a Galactic
  LMXB hosting a neutron star and powered by Roche lobe overflow.

A posteriori we searched for periodic dips in all pointed
observations of GX 13+1. \cite{Trigo_2012} suggested the presence of
possible dipping states in EXOSAT \citep[see ][]{Stella_85}, RXTE/PCA
\citep[see ][]{Schnerr_03}, and XMM/Epic-pn observations \citep[see
][]{Trigo_2012}. We have analysed these observations finding that
these dipping states are not related with the orbital phase. We believe
that the periodic  dips shown in this work are connected to the
periodic motion of the hot spot at the outer accretion disc caused by
the impact of the stream of matter from the companion star. On the
other hand, the dipping state in GX 13+1 observed by \cite{Trigo_2012}
in some XMM observations and the bi-modal behaviour of the hardness ratio
with respect to the intensity of the source observed by \cite{Schnerr_03}
in the RXTE/PCA observations P30050-01 and P30051-01 and by
\cite{Stella_85} in the EXOSAT observation taken on 1983 September 22,
may be caused by partial absorption in the large outflow of  matter
from the inner region of the system detected by several authors \citep[see
e.g.][]{Ueda_01, Ueda_04, Trigo_2012}.

Finally, we also tried to estimate the orbital period derivative and
we reported the quadratic ephemeris of the source. We find 
 only a loose constraint on the period derivative of $8(37) \times 10^{-8}$
s/s, with the error at 68\% confidence level. The large
associated uncertainty to the quadratic term does not allow us to draw
a firm conclusion, and further observations are needed to constrain
this result.

\section*{Acknowledgements}
This research has made use of MAXI data provided by RIKEN, JAXA and
the MAXI team. AR Gratefully acknowledges Sardinia Regional Government
for the financial support (P.O.R. Sardegna F.S.E. Operational
Programme of the Autonomous Region of Sardinia, European Social Fund
2007-2013 - Axis IV Human Resources, Objective l.3, Line of Activity
l.3.1 “Avviso di chiamata per il finanziamento di Assegni di
Ricerca”). Work in Cagliari was partially funded by the Regione
Autonoma della Sardegna through POR-FSE Sardegna 2007-2013,
L.R. 7/2007, “Progetti di ricerca di base e orientata”, Project N°
CRP-60529

\bibliographystyle{aa} 
\bibliography{citations}
\end{document}